# Laser absorption spectroscopy studies to characterize Cs oven performances for the negative ion source SPIDER

**M. Barbisan,**[a] **S. Cristofaro,**[b] **L. Zampieri**[c]**, R. Pasqualotto**[a] **and A. Rizzolo**[a]

[a] *Consorzio RFX (CNR, ENEA, INFN, University of Padova, Acciaierie Venete SpA),*
 *C.so Stati Uniti 4, 35127 Padova, Italy,*

[b] *Max-Planck-Institut für Plasmaphysik,*
 *Boltzmannstr.2, 85748 Garching, Germany*

[c] *Università degli Studi di Padova,*
 *Via 8 Febbraio, 2 - 35122 Padova*
 *E-mail*: `marco.barbisan@igi.cnr.it`

ABSTRACT: The SPIDER $H^-/D^-$ ion source is currently in operation in the Neutral Beam Test facility (NBTF) at Consorzio RFX (Padova, Italy) to prove the possibility of generating up to 40 A of negative ions, with a maximum extracted current density of 350 $A/m^2$ (H)/285 $A/m^2$ (D) and a fraction of co-extracted electrons not greater than 0.5 (H)/1 (D). These performances are required for the realization of the ITER Neutral Beam Injector (NBI), which should deliver 16.7 MW to the plasma by means of negative ions accelerated up to 1 MeV and neutralized before being injected into the ITER tokamak. In order to obtain such high extracted current densities and low co-extracted electron fractions it is necessary to lower the work function of the surface of the acceleration system grid facing the source; this will be accomplished by coating the surfaces with Cs, routinely evaporated by three ovens. The functionality of the ovens has been tested at the CAesium oven Test Stand (CATS), hosted at NBTF. The test stand is equipped with several diagnostics, among which a Laser Absorption Spectroscopy (LAS) diagnostic. Using a tunable laser diode, the LAS diagnostic gets the high resolution absorption spectrum of the Cs 852 nm $D_2$ line along a line of sight to measure Cs density at ground state. The paper describes the test stand and the LAS diagnostic, together with the characterization of the ovens to be installed in SPIDER. The paper will also study the systematic density underestimation effect caused by Cs ground state depopulation, as a function of laser intensity and of Cs density, in the perspective of correcting the density evaluation.

# Contents

## 1. Introduction

The full operation of the ITER experiment requires additional heating systems, among which two Neutral Beam Injectors (NBIs) are foreseen. The ITER NBIs should deliver a beam of 16.7 MW, composed by H or D neutral particles accelerated at about 1 MeV/ 0.87 MeV, respectively [1]. The neutral beams will be obtained by gas neutralization of a $H^-/D^-$ beam, extracted from an ICP (Inductively Coupled Plasma) RF source and accelerated by a system of several grids. In order to reach the above mentioned performances, it is required that negative ions are extracted from the source with a current density of 350 $A/m^2$ ($H^-$)/285 $A/m^2$ ($D^-$). Moreover, in order to limit the heat loads on the various components the fraction of co-extracted electrons should not exceed 0.5 ($H^-$)/1 ($D^-$) [2]. This is not feasible when the only volume production of negative ions in the source (electron capture and dissociation of vibrationally excited $H_2/D_2$ molecules) is exploited; rather, surface production reactions are used and boosted by lowering the work function of the surface of the grid facing the negative ion source (the so called Plasma Grid – PG). This will be accomplished by evaporating Cs into the source by means of Cs ovens. It has been proven that, by carefully controlling the thickness of the deposited Cs layer, the work function of the PG surface can be lowered down to 2.2 eV [3], with an increase of the extracted current density of roughly one order of magnitude and also a lowering of the fraction of co-extracted electrons. The drawback of this solution is the potential non uniformity of the Cs deposition on the PG, which would result in spatial non-uniformities of beam current and, correspondingly, of beam divergence, therefore limiting the NBI performances and reliability. The entity of the issue is increased by the large dimensions of the negative ions sources for the ITER NBI (about 1.77 m x 0.87 m [2]); moreover, Cs is redistributed by the source plasma, often asymmetrical because of the drifts induced by the electrostatic and magnetic fields applied inside the plasma [4].

The physical and technological challenges related to the ITER NBIs are being studied at the Neutral Beam Test Facility (NBTF) of Consorzio RFX, in Padua. The facility hosts the prototype of the ITER NBI negative ion source: SPIDER (Source for the Production of Ions o

Deuterium Extracted from an Rf plasma) [2]; the facility will also host MITICA (Megavolt ITer Injector Concept and Advancement), the full prototype of the ITER NBI [5]. In SPIDER three caesium ovens will be installed in vacuum, on the rear surface of the source [6] to guarantee a sufficient and homogeneous distribution of Cs in the source. Two ovens have been tested at the CAesium oven Test Stand (CATS) in NBTF [6] to verify their reliability and functioning and to gain experience on the control of the Cs output. CATS hosts one oven in a vacuum chamber, equipped with several diagnostics; among them, a Laser Absorption Diagnostic (LAS) allows to measure the Line of Sight (LOS) integrated density of Cs, calculated from the absorption spectrum of the Cs $D_2$ line. The high resolution spectra are obtained using a tunable laser diode, whose emission wavelength, can be varied with sub-pm resolution. The LAS diagnostic in CATS is a prototype of a 4 LoSs system which will monitor the vertical distribution of Cs in the volume of the source, thus complementing the other SPIDER source diagnostics [7].

The present paper will firstly introduce the LAS measurement technique, the Cs ovens, CATS and the LAS diagnostic installed in it. The performances of the Cs ovens, in terms of the Cs density measured by the LAS diagnostic, will then be described. At last, the paper will present an extensive characterization of the phenomenon of ground state depopulation, which takes place in case of a too strong laser intensity and leads to underestimate the Cs density. The effectiveness of a correction method for this source of systematic error is tested.

## 2. LAS principle of operation

The core of the LAS diagnostic is a tunable cw laser diode, whose wavelength can be linearly scanned by changing the current flowing in it. Laser output is routed by an optical fiber to a collimator, which shines a light beam through the vacuum chamber inside which the absorbing medium is present. On the opposite side, the unabsorbed photons are collected by a second collimator and sent through an optical fiber to a detector, whose signal is finally digitized. In the case of the LAS diagnostic for SPIDER, the laser is set and modulated to produce wavelength scans at the Cs $D_2$ line (852.1 nm, $6^2P_{3/2} \rightarrow 6^2S_{1/2}$). The spectral width of the employed laser diode (~$10^6$ Hz FWHM) normally allows to partially resolve the hyperfine structure of the line in two peaks, separated by about 21.4 pm, consisting in the groups of transitions F=3$\rightarrow$ F=2,3,4 and F=4$\rightarrow$ F=3,4,5, respectively [8]. In the first approximation that the population of Cs atoms excited to level $i$ by photon absorption is negligible (and so are stimulated and spontaneous emission), the density of Cs (at ground state $k$) $n_k$ is given by equation 2.1:

$$n_k = \frac{8\pi c}{A_{ik} l \lambda_0^4} \frac{g_k}{g_i} \int ln\left[\frac{I(\lambda, 0)}{I(\lambda, l)}\right] d\lambda \qquad (2.1)$$

where $c$ is the speed of light, $A_{ik} = 3.276 \cdot 10^7 s^{-1}$ is the transition probability for spontaneous emission, $l$ is the length of the optical path in the cesiated volume, $\lambda_0 = $ 852.11 nm is the Cs D2 line central wavelength, $g_k = 2$ and $g_i = 4$ are the statistical weights of lower and upper level, $I(\lambda, 0)$ and $I(\lambda, l)$ are the intensities of the laser light at wavelength $\lambda$ before and after passing through the cesiated volume. While $I(\lambda, l)$ is the acquired signal, $I(\lambda, 0)$ is obtained by fitting the baseline of absorption spectra with a polynomial function. Integration of the spectral lines is performed using Gaussian fits applied to the peaks in the logarithm ratio; this helps in case of peak saturation, because of the broad baseline in the emission spectrum of the laser diode [8]. The spectra are converted from the time basis of the laser diode current scans to a wavelength basis using the known wavelength separation between the absorption peaks.

As previously stated, eq. 2.1 is valid only when the population of Cs excited states is negligible. In conditions of too high beam intensity and too low Cs density, the ground state is depopulated and the LAS measurements give an underestimation of the real Cs density. The depopulation effect can be detected by simply checking LAS measurements against variations of the laser beam intensity. Compensating for this source of systematic error is important when it is not possible to use a low enough laser intensity. Ground state depopulation can be modeled considering the effect of inhomogeneous Doppler broadening [9], which takes into account the selective absorption of photons by the only Cs atoms with a suitable speed to receive the photons at the absorption wavelengths. This model is necessary since, even at room temperature, the Doppler broadening of the absorption peaks is roughly one order of magnitude larger than the natural linewidth [10]. In mathematical terms, with inhomogeneous Doppler broadening the absorption coefficient $k_0 = -(dI/dl)/I$ becomes

$$k = \frac{k_0}{\sqrt{1+\beta I}} \quad \text{where} \quad \beta = \frac{2\sigma}{h\nu A_{ik}}, \qquad (2.2)$$

where $\sigma = 2.3 \cdot 10^{-13} m^2$ is the absorption resonant cross section [11], $h$ is the Planck constant and $\nu$ is the electromagnetic wave frequency. Eq. 2.1 can then be rewritten as follows:

$$n_k = \frac{8\pi c}{\lambda_0^4 A_{ik} l} \frac{g_k}{g_i} \int \{f[I(\lambda, 0)] - f[I(\lambda, l)]\} d\lambda, \qquad (2.3)$$

where the function $f$ is defined as

$$f(I) = 2\sqrt{1+\beta I} + ln\left(\frac{\sqrt{1+\beta I}-1}{\sqrt{1+\beta I}+1}\right) \qquad (2.4)$$

To perform the calculation of eq. 2.3, the absorption spectra must be calibrated in intensity.

## 3. Cs ovens and LAS diagnostic setup in CATS

The tested Cs ovens dedicated to SPIDER [6] consist of a Cs reservoir (inner diameter 30 $mm$ x 58 $mm$ height) connected to a duct (inner diameter 8 $mm$), at whose end the nozzle allows the Cs vapors to flow through 6 apertures (diameter 3.5 $mm$). A solenoid valve installed along the duct allows to stop Cs flux from the reservoir and to protect the Cs content of the reservoir whenever there is a risk of contamination with air. Several heating cartridges and a band heater, feedback controlled by five thermocouples, keep reservoir and duct at constant and independent temperatures, to allow Cs evaporation and prevent Cs accumulation in the duct. A 3D model of the ovens is shown in Figure 1. The sketch reports the position of the thermocouples TC2 (reservoir) and TC4 (duct), used to study Cs ovens performances.

In CATS, the tested oven is hosted in a cylindrical vacuum vessel (inner diameter 400 $mm$ x 927 $mm$ length) kept at $10^{-7} \div 10^{-8} mbar$; the vessel is shown in Figure 2a. In addition to LAS and the Surface Ionization Detector (SID) filaments mounted on the oven nozzle (Figure 1), several diagnostics can be installed in CATS: an infrared thermocamera, a quartz microbalance, a Residual Gas Analyzer (RGA) and a second, freely movable SID system [12]. The LAS transmission and reception collimators are mounted on Kodial windows, at the end of $125\ mm$ long pipes, in order to prevent the direct exposure of the windows to the direct flux of Cs. For the same reason, the LoS is $90\ mm$ apart from the nozzle axis. This is well visible in Figure 2b, showing the 3D model of CATS from the top; the exit cones of the Cs flux from the nozzle apertures are shown in red. While in the first campaign with LAS in CATS [10] $l$ was assumed to be the window-window distance ($80\ cm$), recent simulations of Cs distribution inside CATS [13] have shown that Cs density is quite higher in the vacuum vessel cylinder than in the LAS pipes. It was then established to consider only the segment of the LoS in direct view of the nozzle, i.e. $l = 48\ cm$. The values of Cs density obtained under this convention must be examined keeping into consideration that, according to the simulations, the adopted LoS segment comprises about 75 % of the observed Cs atoms.

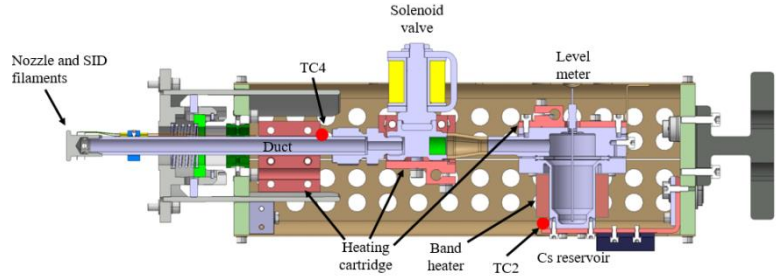

**Figure 1.** 3D model of a Cs oven for SPIDER, view in vertical section.

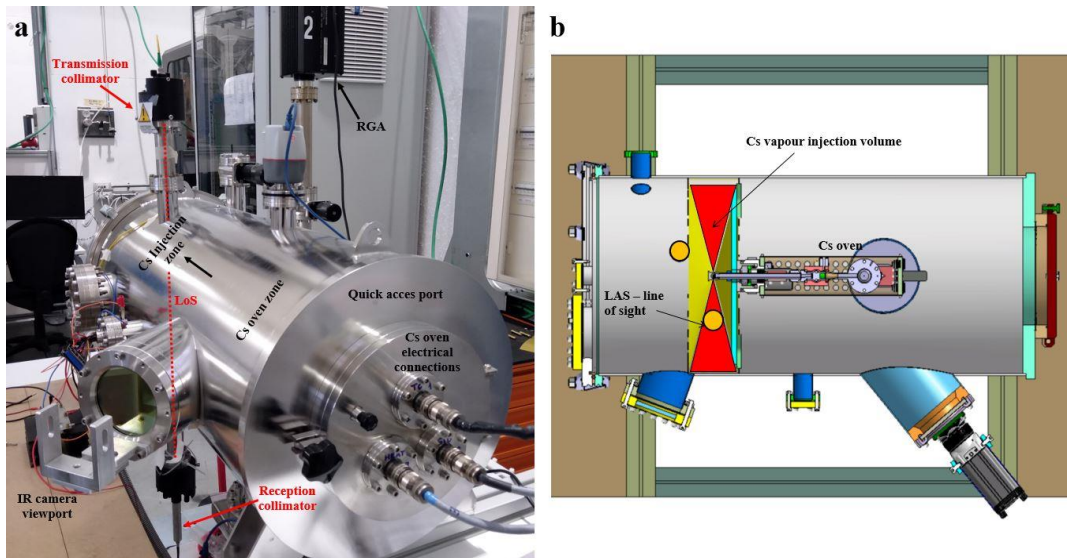

**Figure 2.** a) Picture of the CATS test bed from the side the CS oven is inserted. b) Top view of the 3D model of the CATS test bed.

Regarding the LAS diagnostic hardware, the laser diode is a *Sacher Lasertechnik* DFB-0852-150-TO3, controlled in temperature ($0.06\ nm/K$) and current ($3\ pm/mA$) by a separate device (*S.L.* Pilot PC 500). The laser diode is protected by two optical isolators, one in the laser head and an external one, providing 70 dB isolation from back reflections in total. The laser light is carried through single mode 5μm(core)/125μm(cladding) fibers. In order to study depopulation, the laser power can be reduced by more than 4 orders of magnitude, by means of two voltage controlled fiber attenuators (*Thorlabs* V800A). The fiber leads the light to the transmission collimator, producing a laser beam of 6.5 $mm$ diameter. At the opposite window, the reception optic head collects and conveys light on a $1000\mu m/1100\mu m$ multimode fiber to the detection system. The light exiting the fiber is coupled to the detector by means of an aspheric lens with 10 mm optical aperture and 8 mm focal length. The coupling system also hosts an interference filter to block spurious light (*Andover C.* 064FSX10-12.5, with 10 nm FWHM bandwidth) and a neutral density filter to match light power down to the range of the detection system if needed. The photodiode module, extensively described in [14], provides more than 4 orders of magnitude of dynamic range and a bandwidth from 1.2 kHz to 135 KHz, more than adequate for the typical LAS scan rate of 1 Hz. At last, a *National Instruments* NI PXIe-6259 module digitizes the signals from the detector (0.25 $MS/s$ multiplexed, 16 bit resolution) and generates the modulation signal for the laser (1.25 $MS/s$, 16 bit resolution).

## 4. Characterization of Cs ovens

In 2019 the LAS diagnostic was active during the test of the first two SPIDER Cs ovens. Figure 3a shows the Cs density measurements obtained in steady state conditions in CATS, as functions of oven reservoir temperature. The data of oven 1 and 2 are shown with circles and squares, respectively. Blue and orange symbols indicate data taken with the oven duct temperature set at $200°C$ and $250°C$, respectively. Data shown with void circles refer to the second campaign on oven 1; between the two campaigns, the oven was ventilated in nitrogen, the valve was then closed and the oven was extracted and left in air for 3 weeks. The data represented with void squares refer to the second campaign of oven 2; in this case, in between the campaigns the oven was exposed for 10 days to air, with the valve closed and the reservoir kept in vacuum. What results from experimental data is that oven 1 and 2, at least in the respective first campaigns, have comparable performances and that the measured Cs density follows a common exponential relation with respect to reservoir temperature. Switching the duct temperature between 200°C and 250°C does not seem to alter this relation. Density measurements of the second campaign of oven 1 are instead more than one order of magnitude lower than measurements taken during the first campaign with similar values of reservoir temperature. This was explained by the subsequent inspection of oven 1: a leakage led to complete oxidation of Cs inside the reservoir. In the case of the second campaign on oven 2, again density measurements were found lower than before but not as much as for oven 1. In this case, no contamination of the reservoir was found and also the SID diagnostics indicated no significant reduction of the Cs flux. The data of the second campaigns should then be studied also in terms of the amount of Cs which was accumulated on the inner walls of the vacuum vessel and in the duct in the first campaign, and then lost during the exposure to air.
The dependence of density measurements on Cs accumulated on the vessel walls and in the duct can be better realized from Figure 3b, showing the time trend of density measurements, right after valve aperture, at same duct and reservoir temperatures ($200°C$ and $110°C$, respectively). The data of the black curve belong to a day at the beginning of the first experimental campaign of oven 1 (30/05/2019); the data of the red curve were taken after one week of operation (05/06/2019). In the second case, the growth of Cs density is faster and also the final steady state level is higher (about $5.5 \cdot 10^{15} m^{-3}$ against $3.0 \cdot 10^{15} m^{-3}$ in the first case). This gives a reproducibility estimate of Cs ovens performances in CATS.

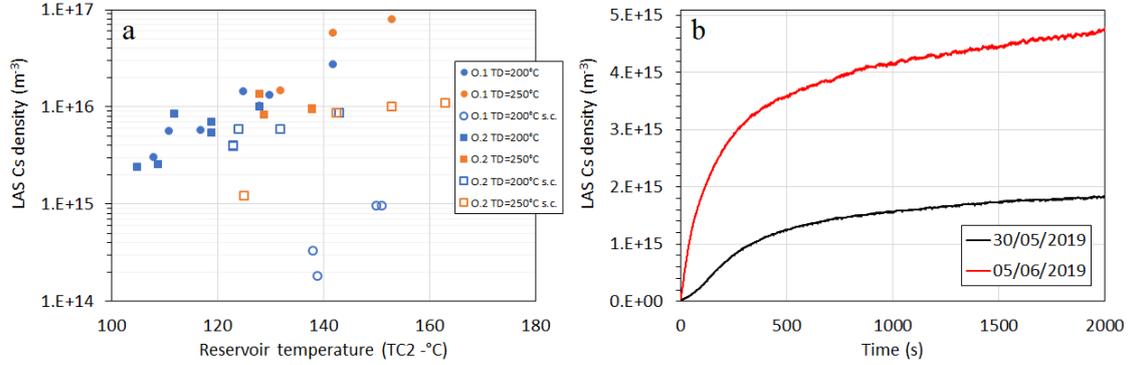

**Figure 3.** a) Cs density measured by LAS in CATS, as function of oven reservoir and duct temperatures. Circles and squares indicate data points of oven 1 and 2, respectively. Blue and orange colors differentiate data taken with duct temperature at $200°C$ and $250°C$, respectively. Void symbols indicate data from the second experimental campaign of each oven. b) Temporal evolution of Cs density measured by LAS in CATS with oven 1, measured at the beginning (30/05/2019) of the experimental campaign (black curve) and after 1 week of operation (05/06/2019, red curve). Time 0 indicates the moment of oven valve opening.

## 5. Correction of depopulation effects

As explained in sec. 2, with excessive light beam intensity the Cs ground state is depopulated and as consequence $n_k$ in eq. 2.1 is underestimated. A first attempt to correct the measurements by applying eqs. 2.3 and 2.4 was already performed, with partial success [10]. The effectiveness of the depopulation correction is now studied also as a function of the Cs density itself, in a range of laser intensity from $0.03\ W/m^2$ to $20\ W/m^2$, exploiting the double attenuator in series to the laser. LAS diagnostic measurements were performed with Cs density in CATS at several steady state levels, by varying the laser beam intensity. When reaching a stable Cs evaporation was too long (several hours can be required), Cs density measurements right before and after the laser intensity scans were fitted with a 1° or 3° degree polynomial to correct the measurements in the scan. Figure 4a shows the results of these scans, differentiated by color; each point represents the average value of the density measurements performed at a given laser intensity. Void circles indicate Cs density values as given by eq. 2.1, while the values given by the depopulation corrected method of eqs. 2.3 and 2.4 are plotted with full circles. What results is that ground state depopulation, as expected, is more relevant with lower Cs density; detectable underestimation of density appears between $0.1\ W/m^2$ and $1\ W/m^2$. It results also that, at laser intensity values up to $20\ W/m^2$, the depopulation correction method is effective in keeping Cs density sufficiently stable with respect to laser intensity. At last, Cs density is underestimated at laser intensity below $0.1\ W/m^2$. This should not be attributed to ground state depopulation, but to the distorsion of the absorption spectra baseline, that is introduced by the attenuators when too high level of attenuation is requested.

From laser intensity scans not just the Cs density but also the ratio between the absorption peaks were considered. From the previous study [10] it emerged that, with increasing laser intensity, the $F = 3$ ground state sublevel is more depopulated than the $F = 4$ one. This is a limit for the measurement correction method, which does not consider this phenomenon. For the laser intensity scans shown in Figure 4a, Figure 4b shows the corresponding average values of the ratio between the absorption peak integrals at lower and higher wavelength, as function of laser intensity. Only the values obtained by using the uncorrected method are shown, since the depopulation correction does not introduce appreciable variations in this quantity. What results is that most of the scans exhibit similar trends for the peaks ratio; the only two scans in which the ratio is kept more constant with respect to laser intensity are those at highest Cs density.

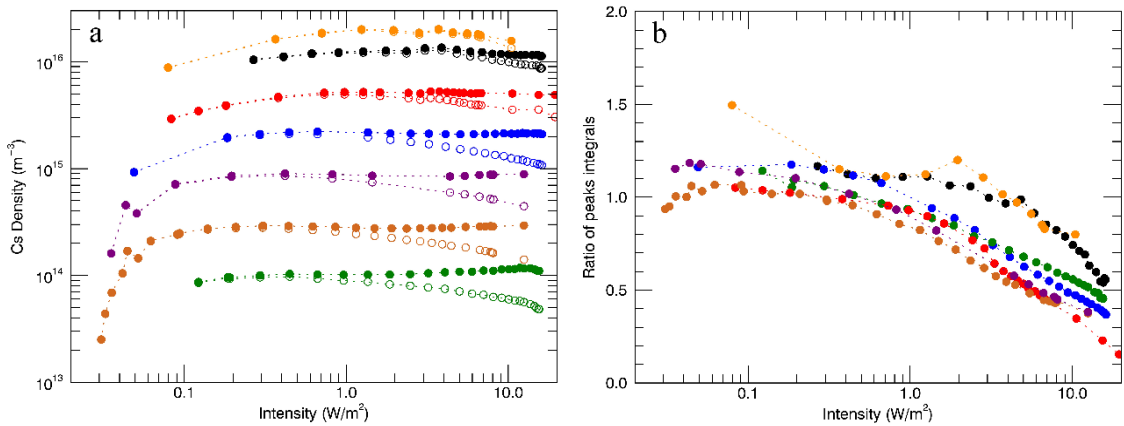

**Figure 4.** a) Sets of Cs density measurements in steady conditions by LAS in CATS, as function of laser beam intensity. Void circles indicate data calculated according to eq. 2.1, while full circles of the same color were obtained with depopulation correction. b) Ratio between the integrals of the absorption peaks at lower and higher wavelength, as a function of laser beam intensity. The data series are associated by color to the analogous points in figure a). The values were calculated by applying eq. 2.1.

## 6. Conclusions

Controlling the Cs distribution in the SPIDER negative ion source is a key point to extract a sufficiently high and uniform beam current from it. Besides monitoring the dynamic of Cs density in the source, it is necessary to properly manage the Cs flux from the ovens installed in SPIDER. The LAS diagnostic is able to provide valuable information for both purposes. In the CATS test bed, the LAS measurements proved that the two tested ovens have similar performances and detected the contamination of one oven's Cs content due to a leakage after external exposure to air. It was also possible to estimate the influence on Cs flux of continuous accumulation of Cs with time; it is still not possible however to distinguish between the contribution of Cs oven duct, between nozzle and valve, from the inner walls of CATS vacuum vessel. Regarding the accuracy of LAS measurements, the previous characterization of ground state depopulation [10] has been extended, verifying the dependence of this effect on both laser intensity and Cs density. The correction method illustrated in sec. 2 proved to be effective at least for laser intensity values up to $20 \ W/m^2$. The experimentation on SPIDER Cs ovens and on LAS measurements will proceed by producing a hydrogen plasma in CATS and by checking its influence on Cs oven operation and on diagnostics measurements.

**Acknowledgments**

The author thanks M. Fadone of Consorzio RFX for the study on Cs density homogeneity along the LoS path. The work leading to this publication has been funded partially by Fusion for Energy under the contract n°F4E-OFC-531-1. This publication reflects the views only of the authors, and F4E cannot be held responsible for any use which may be made of the information contained therein. The views and opinions expressed herein do not necessarily reflect those of the ITER Organization.